\documentclass[twocolumn]{aastex62}
\usepackage{natbib}
\usepackage{threeparttable}
\usepackage{amssymb}
\usepackage{graphicx}
\usepackage{CJKutf8}
\usepackage{textcomp}
\usepackage{rotating}

\newcommand{\dn}{$\mbox{D}_n4000$}
\newcommand{\EWHd}{EW(H$\delta$)}

\shorttitle{Stellar Populations at $z\sim1$}
\shortauthors{Wu et al.}

\begin{document}

\title{Stellar Populations of over one thousand $z\sim0.8$ Galaxies from LEGA-C: Ages and Star Formation Histories from D$_n$4000 and H$\delta$}

\author{Po-Feng Wu \begin{CJK*}{UTF8}{bkai}(吳柏鋒)\end{CJK*}}
\affiliation{Max-Planck-Institut f\"{u}r Astronomie, K\"{o}nigstuhl 17, D-69117, Heidelberg, Germany}

\author{Arjen van der Wel}
\affiliation{Max-Planck-Institut f\"{u}r Astronomie, K\"{o}nigstuhl 17, D-69117, Heidelberg, Germany}
\affiliation{Sterrenkundig Observatorium, Universiteit Gent, Krijgslaan 281 S9, B-9000 Gent, Belgium}

\author{Anna Gallazzi}
\affiliation{INAF-Osservatorio Astrofisico di Arcetri, Largo Enrico, Fermi 5, I-50125 Firenze, Italy}

\author{Rachel Bezanson}
\affiliation{University of Pittsburgh, Department of Physics and Astronomy, 100 Allen Hall, 3941 O'Hara St, Pittsburgh PA 15260, USA}

\author{Camilla Pacifici}
\affiliation{Astrophysics Science Division, Goddard Space Flight Center, Code 665, Greenbelt, MD 20771, USA}
\affiliation{Space Telescope Science Institute, 3700 San Martin Drive, Baltimore, MD 21218, USA}

\author{Caroline Straatman}
\affiliation{Max-Planck-Institut f\"{u}r Astronomie, K\"{o}nigstuhl 17, D-69117, Heidelberg, Germany}

\author{Marijn Franx}
\affiliation{Leiden Observatory, Leiden University, PO Box 9513, 2300 RA Leiden, The Netherlands}

\author{Ivana Bari\v{s}i\'{c}}
\affiliation{Max-Planck-Institut f\"{u}r Astronomie, K\"{o}nigstuhl 17, D-69117, Heidelberg, Germany}

\author{Eric F. Bell}
\affiliation{Department of Astronomy, University of Michigan, 1085 South University Avenue, Ann Arbor, MI 48109-1107, USA}

\author{Gabriel B. Brammer}
\affiliation{Space Telescope Science Institute, 3700 San Martin Drive, Baltimore, MD 21218, USA}

\author{Joao Calhau}
\affiliation{Physics Department, Lancaster University, Lancaster LA1 4YB, UK}

\author{Priscilla Chauke}
\affiliation{Max-Planck-Institut f\"{u}r Astronomie, K\"{o}nigstuhl 17, D-69117, Heidelberg, Germany}

\author{Josha van Houdt}
\affiliation{Max-Planck-Institut f\"{u}r Astronomie, K\"{o}nigstuhl 17, D-69117, Heidelberg, Germany}

\author{Michael V. Maseda}
\affiliation{Leiden Observatory, Leiden University, PO Box 9513, 2300 RA Leiden, The Netherlands}

\author{Adam Muzzin}
\affiliation{Department of Physics and Astronomy, York University, 4700 Keele St., Toronto, Ontario, M3J 1P3, Canada}

\author{Hans-Walter Rix}
\affiliation{Max-Planck-Institut f\"{u}r Astronomie, K\"{o}nigstuhl 17, D-69117, Heidelberg, Germany}

\author{David Sobral}
\affiliation{Physics Department, Lancaster University, Lancaster LA1 4YB, UK}

\author{Justin Spilker}
\affiliation{Department of Astronomy, University of Texas at Austin, 2515 Speedway Stop C1400, Austin, TX 78712, USA}

\author{Jesse van de Sande}
\affiliation{Sydney Institute for Astronomy, School of Physics, University of Sydney, NSW 2006, Australia}

\author{Pieter van Dokkum}
\affiliation{Astronomy Department, Yale University, New Haven, CT 06511, USA}

\author{Vivienne Wild}
\affiliation{School of Physics and Astronomy, University of St Andrews, North Haugh, St Andrews, KY16 9SS, U.K.}

\correspondingauthor{Po-Feng Wu}
\email{pofeng@mpia.de}

\begin{abstract}

Drawing from the LEGA-C dataset, we present the spectroscopic view of the stellar population across a large volume- and mass-selected sample of galaxies at large lookback time. We measure the 4000\AA\ break (D$_n$4000) and Balmer absorption line strengths (probed by H$\delta$) from 1019 high-quality spectra of $z=0.6 - 1.0$ galaxies with $M_\ast = 2 \times 10^{10} M_\odot - 3 \times 10^{11} M_\odot$. Our analysis serves as a first illustration of the power of high-resolution, high-S/N continuum spectroscopy at intermediate redshifts as a qualitatively new tool to constrain galaxy formation models. The observed D$_n$4000-EW(H$\delta$) distribution of our sample overlaps with the distribution traced by present-day galaxies, but $z\sim 0.8$ galaxies populate that locus in a fundamentally different manner. While old galaxies dominate the present-day population at all stellar masses $> 2\times10^{10} M_\odot$, we see a bimodal D$_n$4000-EW(H$\delta$) distribution at $z\sim0.8$, implying a bimodal light-weighted age distribution. The light-weighted age depends strongly on stellar mass, with the most massive galaxies $>1\times10^{11}M_\odot$ being almost all older than 2 Gyr. At the same time we estimate that galaxies in this high mass range are only $\sim3$~Gyr younger than their $z\sim0.1$ counterparts, at odd with pure passive evolution given a difference in lookback time of $>5$~Gyr; younger galaxies must grow to $>10^{11}M_\odot$ in the meantime, and/or small amounts of young stars must keep the light-weighted ages young. Star-forming galaxies at $z\sim0.8$ have stronger H$\delta$ absorption than present-day galaxies with the same D$_n$4000, implying larger short-term variations in star-formation activity.

\end{abstract}


\section{Introduction}

The Sloan Digital Sky Survey \citep[SDSS;][]{yor00} produced one of the most valuable legacy datasets for galaxy evolution studies. From the strengths and shapes of spectral lines, the SDSS spectra provide diagnostics for fundamental physical properties of individual galaxies: ages and metal content of stellar populations, star-formation rates (SFRs), metallicity in the inter-stellar medium (ISM), and internal dynamics. Furthermore, with hundreds of thousands of spectra, the SDSS had characterized various galactic scaling relations \citep[][to name a few]{bri04,tre04,gal05}. This information shaped our understanding of the formation of galaxies.   

Despite its tremendous success, the SDSS is mainly confined to the nearby Universe. The median redshift of the SDSS spectroscopic main sample is $z\sim0.1$, which corresponds to $\sim 1$ Gyr of look-back time \citep{str02}. 
On the other hand, deep wide-field optical and near infrared (NIR) imaging surveys have pushed the census of galaxy population to $z\sim4$ \citep{mar09,ilb13,muz13}. From photometric studies, we have constructed the growth history of the stellar mass density as a function of cosmic time. We know that $\sim90\%$ of stars form after $z\sim2$ and about half of stars formed since $z\sim1$ \citep{rud03,muz13,mad14}. The relative abundance of quiescent and star-forming galaxies also evolves with cosmic time. At low redshifts, massive galaxies are mainly quiescent, while at $z\gtrsim2$, star-forming galaxies become the dominant population at all masses \citep{ilb10,muz13}. These observations show that the stellar population in high redshift galaxies are very different from local galaxies. However, we have not yet understood the processes driving the assembly of stellar masses and shaping the star-forming properties of galaxies throughout cosmic time.

Spectroscopic redshift surveys, such as DEEP2 \citep{new13}, zCOSMOS \citep{lil07}, VVDS \citep{lef13}, or VIPERS \citep{guz14,gar14}, have gathered tens of thousands of galaxy spectra using multi-object spectrographs on 8-10~m-class telescopes, thereby pushing the spectroscopic census of galaxy population to $z\sim1$ and beyond. 
In order to obtain a large number of spectra, these surveys need to compromise on the signal-to-noise ratio or spectral resolution in exchange for sample sizes. They provide a profound resource for studying the star-formation and ISM properties through emission lines. However, the quality of these spectra is usually not good enough to constrain the ages and metallicities of stars in individual galaxies through the stellar continuum. 
So far, our understanding of stellar populations of galaxies at $z \sim 1$ and beyond only comes from studies with sample sizes of a few dozens, mainly massive and quiescent galaxies \citep{kel01,tre05,vdw05,jor13,vds13,gal14,cho14,bel15,ono15}. 
This is far from representative of the galaxy population at that epoch. 

To achieve both the depth and sample size required for characterizing the stellar content in the early Universe, we carry out the Large Early Galaxy Astrophysics Census (LEGA-C) survey \citep{vdw16}. 
The LEGA-C survey will obtain $\sim 3000$ $K_s$-band-selected spectra at $z\sim1$ with typical signal-to-noise ratio (S/N) of 20 \AA$^{-1}$. The quality of the spectra allows us to characterize the stellar populations of individual galaxies and galaxies as a population, akin to what has been achieved by the SDSS \citep{kau03a,bri04,gal05}.

In this paper, we present measurements of two age-sensitive absorption line indices, the equivalent width of H$\delta$ absorption [EW(H$\delta$)] and D$_n$4000 index, of 1019 galaxies selected from the LEGA-C survey. For a simple stellar population (SSP), the D$_n$4000 index increases monotonically with time. On the other hand, the EW(H$\delta$) increases rapidly in the first few hundreds Myrs when the O- and B-type stars fade and the A-type stars dominate the spectrum. The EW(H$\delta$) then decreases afterwards when A-type stars also fade. For a composite stellar population, the peak strength of the H$\delta$ absorption depends on whether the star-formation rate varies rapidly or changes smoothly. These two spectral features have been extensively used as diagnostics for the ages of the stellar population and to discern recent star-formation histories \citep{kau03a,leb06,kau14,mal16}. 

In the local Universe, \citet{kau03b} showed that both the D$_n$4000 and EW(H$\delta$) of galaxies exhibit bimodal distributions, suggesting a bimodality in the light-weighted stellar ages. On average, lower-mass galaxies have smaller D$_n$4000 and larger EW(H$\delta$), which indicate younger stellar populations. Furthermore, for star-forming galaxies, low-mass galaxies have stronger H$\delta$ absorption and the scatter of EW(H$\delta$) at fixed D$_n$4000 is larger than massive star-forming galaxies. These features suggest that low-mass star-forming galaxies have more bursty star-formation histories (SFHs) \citep{kau03a,kau14}. 

Recent spectroscopic surveys has pushed the census on the stellar ages of galaxies to higher redshifts. Similar to galaxies in the local Universe, the D$_n$4000 of galaxies varies with the stellar mass and the bimodal distribution is in place up to $z\sim1$ \citep{ver08,hai17}. Studies on the EW(H$\delta$) is limited, targeting mainly on quiescent galaxies and through stacked spectra \citep{siu17}. 
Because of the typically low S/N and/or low spectral resolution of high-redshift spectroscopic surveys, the uncertainty of EW(H$\delta$) measurements on individual galaxies is too large and the emission line infilling cannot be estimated, preventing accurate constraints on recent star-formation activity.

In this paper we show that the individual LEGA-C spectrum contains precise age information for both star-forming and quiescent galaxies. With over 1000 galaxies, we are able to describe the average age and the patterns of recent star-formation activities at a look-back time of$\sim 7$~Gyrs. We describe the galaxy sample and the quality of the spectral index measurements in Section~2. In Section~3, we present the distribution of D$_n$4000 and EW(H$\delta$) at $z\simeq0.8$ and the comparison to SDSS galaxies at $z\simeq0.1$. We discuss the implications of our results in Section~4 and Section~5. We summarize the paper and point out future directions in Section~6. 

\section{Data and Analysis}
\label{sec:data}

\subsection{The LEGA-C Sample at $z\simeq0.8$}

\begin{figure*}
	\includegraphics[width=\textwidth]{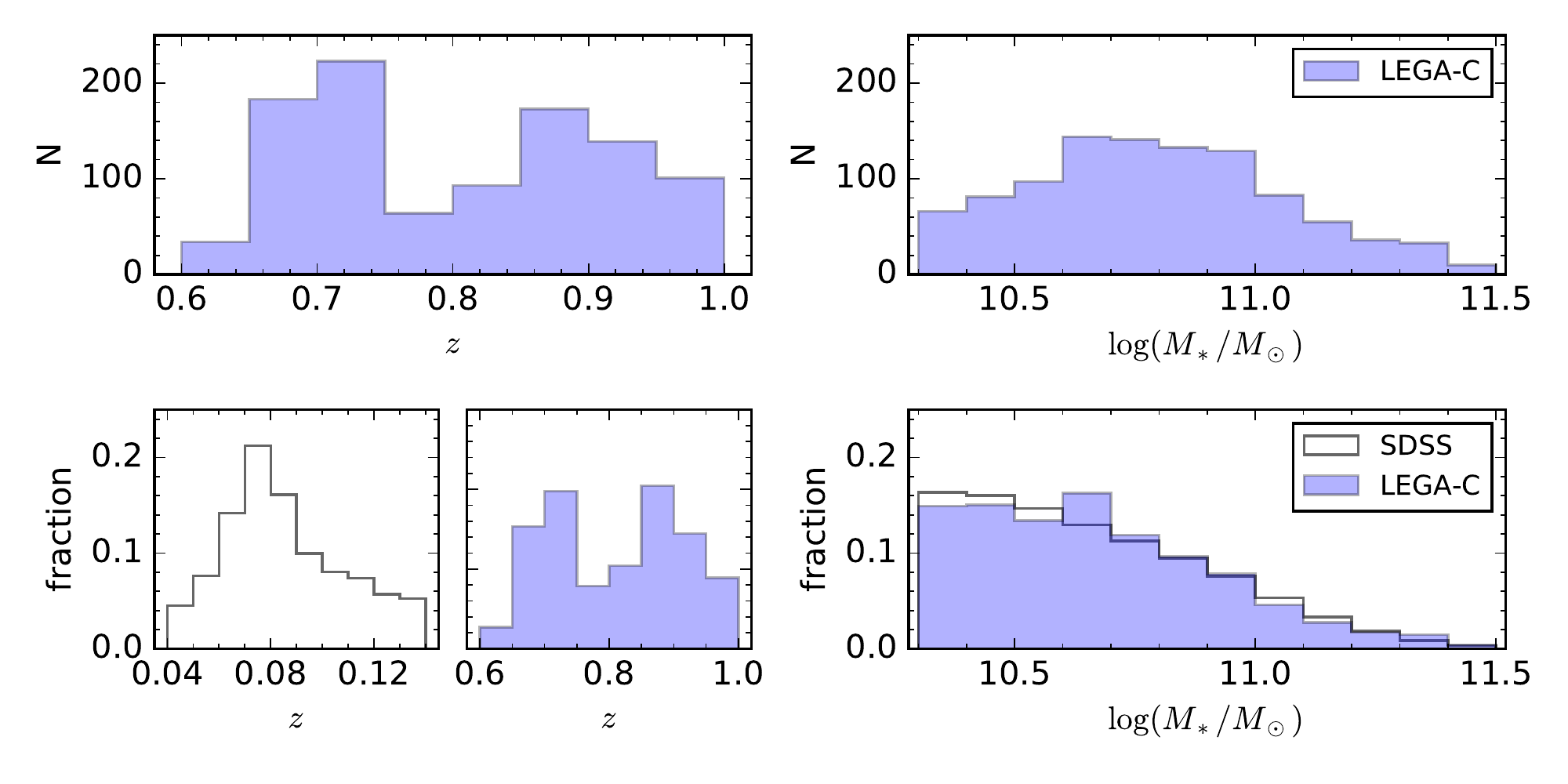}
	\caption{The distributions of redshifts and stellar masses of the LEGA-C and the SDSS sample. \textit{Upper panels:} The histograms of the LEGA-C sample. \textit{Lower panels:} The distributions of the LEGA-C (blue) and SDSS (white) samples with the completeness correction (see Section~\ref{sec:data}).  }
	\label{fig:mzs}
\end{figure*}

The LEGA-C survey is a 4-year survey using the Visible Multi-Object Spectrograph \citep[VIMOS;][]{lef03} mounted on the 8~m Very Large Telescope to obtain rest-frame optical spectra of $\sim 3000$ $K_s$-band selected galaxies mainly at $0.6 \leq z \leq 1.0$. Each galaxy receives $\sim 20$~hrs of integration at a spectral resolution of $R\sim3500$. The typical continuum signal-to-noise ratio (S/N) is 20 \AA$^{-1}$

This study is based on the first two years of data of the LEGA-C survey. The primary sample of the LEGA-C survey consists of those galaxies brighter than $K_s = 20.7 - 7.5 \times \log((1+z)/1.8)$ and with redshifts $0.6 \leq z \leq 1.0$ \citep{vdw16}. From the LEGA-C primary sample, we then select galaxies with stellar mass $10.3 \leq \log(M_\ast/M_\odot) \leq 11.5$ to make a mass-limited sample. The lower mass limit ensures that the $K_s$-band magnitude-limit of the LEGA-C survey does not introduce a strong bias against red galaxies at the low mass end. We also require that the spectra cover the wavelength range for measuring the D$_n$4000 and EW(H$\delta$). We then exclude galaxies detected in X-ray, whose spectra are usually contaminated by the AGN.

There are in total 1050 galaxies fulfill the redshift and stellar mass criteria. We then also require a minimum median $S/N=5$\AA$^{-1}$ between rest-frame wavelength 4000~\AA\ and 4300~\AA. This $S/N$ cut exclude 31 galaxies, $\sim 3\%$ of the sample. The spectral indices of these low-S/N spectra are mostly unphysical, therefore, we decide to exclude them from the sample. 
The majority of these galaxies are bright enough in $K_s$-band to be included in the survey, but have red colors and faint optical magnitudes, resulting in low S/N spectra. They tend to have axis ratios $<0.5$. These galaxies are likely edge-on galaxies whose optical light is attenuated due to the inclination. We have also included these galaxies and repeated our analysis in the paper, the results are not affected. The final sample contains 1019 galaxies from the 1550 galaxies.

We derive galaxy stellar masses by fitting the observed multi-wavelength spectral energy distributions (SEDs) from the UltraVISTA catalog \citep{muz13} using the FAST code \citep{kri09}. The SED templates are from the \citet{bc03} stellar population synthesis models with exponentially declining star-formation rates. We adopt a \citet{cha03} initial mass function (IMF) and the \citet{cal00} dust extinction law. The SFRs are estimated from the UV and IR luminosities, following the prescription of \citet{whi12}. 
The distribution of redshifts and stellar masses of the sample is shown in Figure~\ref{fig:mzs}.

Every galaxy has a volume completeness correction that consists of the traditional $V_{max}$ correction and a survey completeness correction. Both corrections are well understood, as the $K_s$-band flux is the only factor that determines the probability that a galaxy is part of the LEGA-C survey \citep{vdw16}. We refer to the forthcoming Data Release paper for details (Straatman et al. in prep). We apply the completeness correction when comparing the LEGA-C sample to the completeness-corrected SDSS sample (see Section~\ref{sec:sdss}).

\subsection{Measuring D$_n$4000 and EW(H$\delta$) from LEGA-C spectra}

\begin{figure*}
	\includegraphics[width=\textwidth]{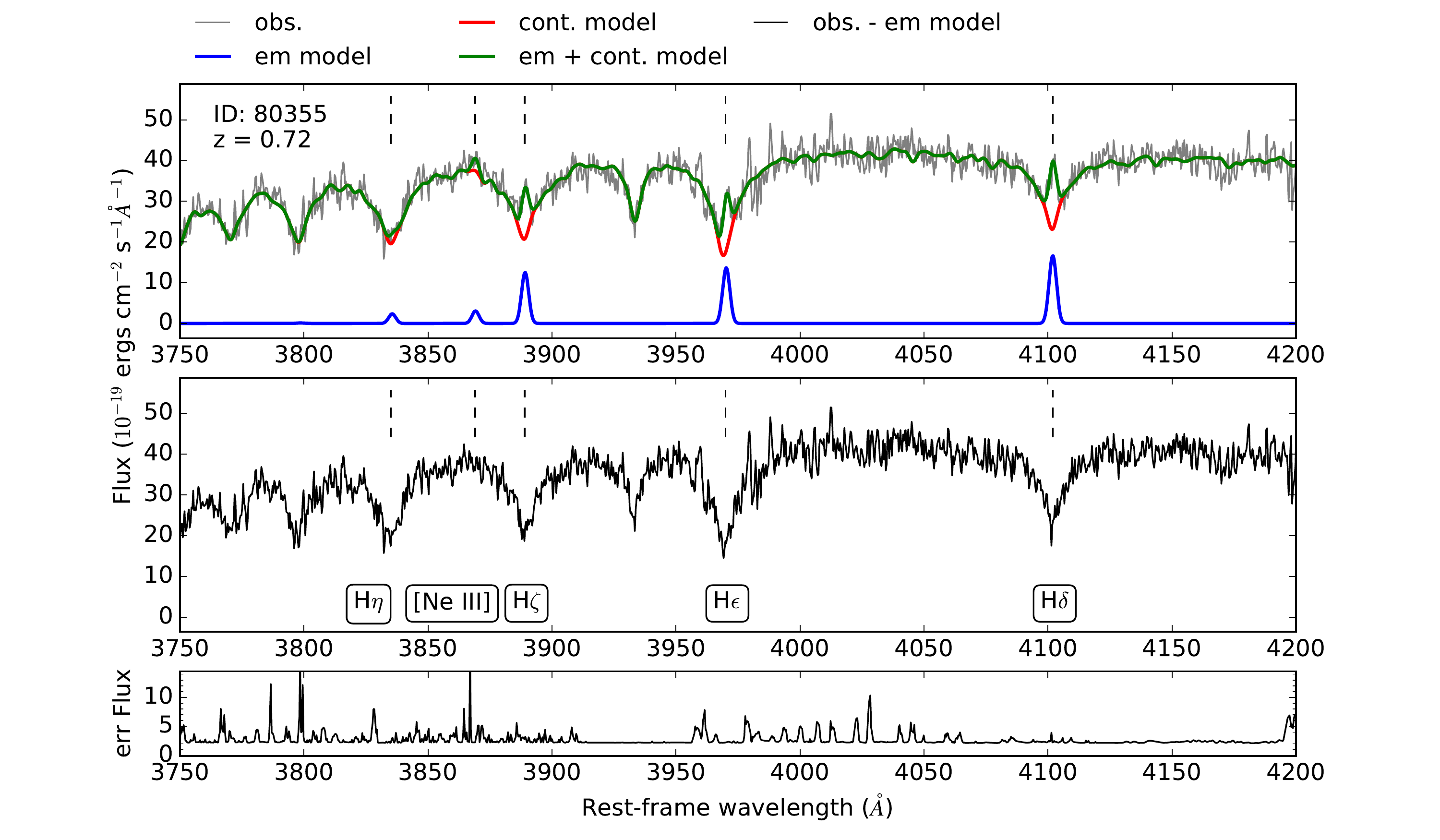}
	\caption{An example of the LEGA-C spectrum and the best-fit model. The gray line in the upper panel shows the observed spectrum near 4000\AA. The stellar continuum (red) and the line emission (blue) are fit simultaneously. Important spectral lines are labeled with vertical dashed lines. The green line is the best-fit model (continuum plus line emission). We then subtract the best-fit emission line model from the observed spectrum (middle panel). The EW(H$\delta$) and D$_n$4000 are measured from the emission-line-subtracted spectrum. The bottom panel shows the uncertainty. }
	\label{fig:model_spec}
\end{figure*}

\begin{figure*}
	\includegraphics[width=\textwidth]{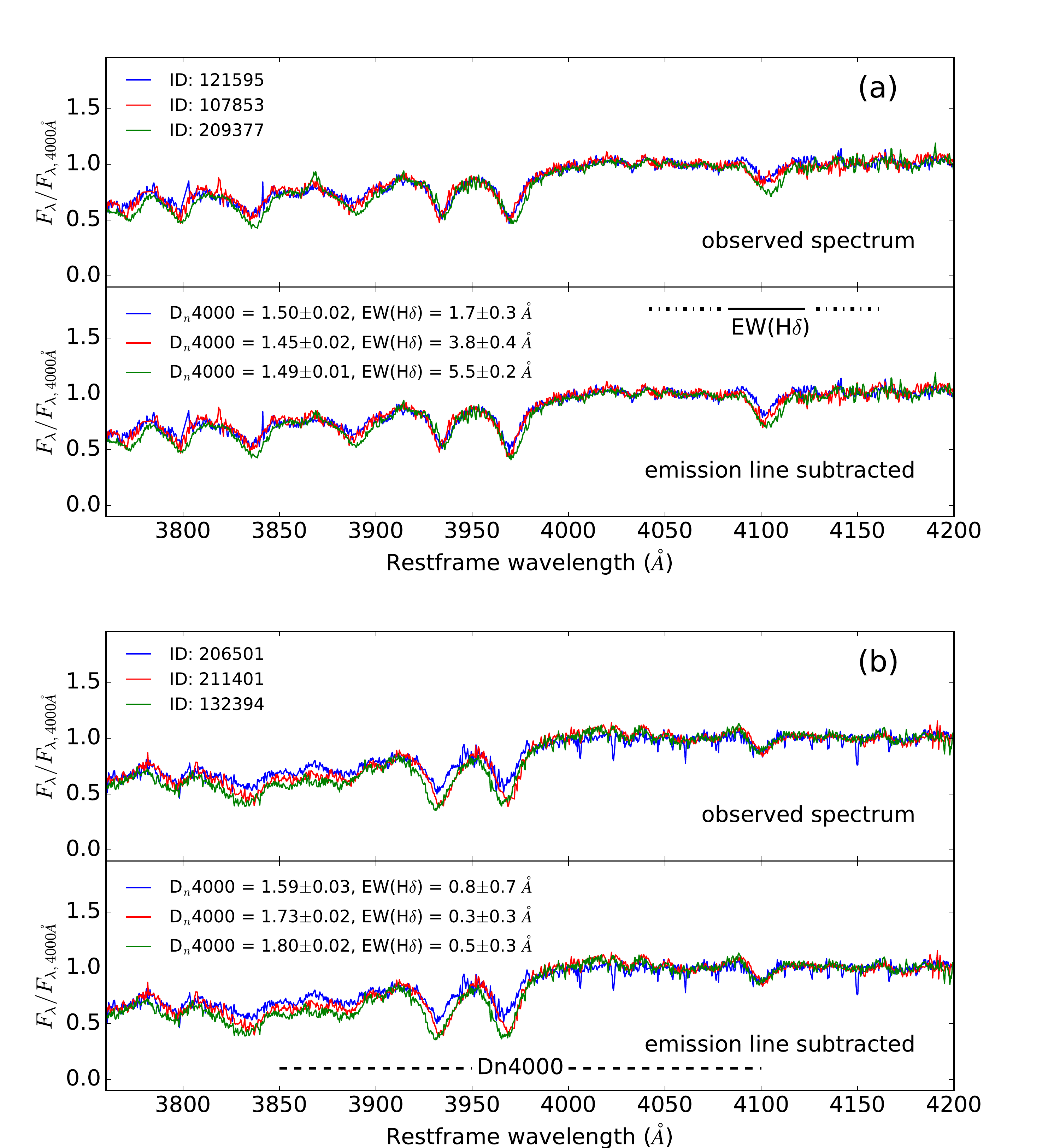}
	\caption{The comparison among the spectra of galaxies with different ages. (a) Three galaxies with comparable D$_n$4000 but different EW(H$\delta$). The SSP-equivalent ages of the three galaxies are between $\sim1-2$~Gyrs. (b) Three galaxies with comparable EW(H$\delta$) but different D$_n$4000. The SSP-equivalent ages of the three galaxies are between $\sim2-3$~Gyrs. The shapes of spectra of different stellar ages can be clearly identified through visual inspection. For each spectrum, the flux is normalized relative to the flux around 4000\AA. The upper and the bottom panel shows spectra before and after subtracting emission line models, respectively. The dashed-dotted lines and the solid lines above the spectra show the bands for measuring the EW(H$\delta$). The dashed lines in the bottom indicate the wavelength ranges of the blue and the red bands for computing the D$_n$4000. Narrow spikes in the spectra are due to imperfect sky subtraction at the locations of bright atmospheric emission lines. }
	\label{fig:DnC}
\end{figure*}

In this paper we measure two stellar absorption line indices: the 4000\AA\ break, D$_n$4000, and the equivalent width of the Balmer absorption, EW(H$\delta$). To separate the stellar continuum from the ionized gas emission, we model the observed spectrum using the Penalized Pixel-Fitting (pPXF) method \citep{cap04} with the updated Python routines \citep{cap17}. Each galaxy spectrum is fit by a combination of two templates representing the stellar and the gas emission. The stellar template is a linear, optimal non-negative combination of \citet{vaz99} SSP models with the Medium resolution INT Library of Empirical Spectra \citep[MILES;][]{san06} empirical stellar spectra and \citet{gir00} isochrones. All emission lines are fit as a single kinematic component, i.e., with the same velocity and velocity dispersion, but the strength of each line is a free parameter. We refer to Bezanson et al. (2017) for the detailed fitting process and Fig.~\ref{fig:model_spec} for an example. 

We adopt the definition of the D$_n$4000 in \citet{bal99} and the H$\delta_a$ index in \citet{wor97} as our EW(H$\delta$). Both indices are measured from emission-line-subtracted spectra. 
The emission line subtraction has little effect on D$_n$4000 but is important for EW(H$\delta$). Our visual inspection suggests that the fit captures weak emission line infilling well. Using 25 galaxies observed twice by the LEGA-C survey, we estimate the uncertainty on our emission line strength measurements. We estimate the typical uncertainties of our final D$_n$4000 and EW(H$\delta$) measurements to be $\sim 0.03$ and $\sim0.4$\AA, respectively.

In Fig.~\ref{fig:DnC}a, we show three galaxies with SSP ages of $\sim1-2$~Gyrs (see Section~\ref{sec:result}). Our spectra clearly differentiate the evolution of the H$\delta$ strength within $\sim1$~Gyrs. Fig.~\ref{fig:DnC}b shows three galaxies with older SSP ages of $\sim 2-3$~Gyrs. The different shapes of the continua can be easily identified by visual inspection and quantified by the D$_n$4000 index.

\subsection{SDSS sample at $z\sim0.1$}
\label{sec:sdss}
From the SDSS DR7 \citep{aba09}, we first select galaxies from a narrow redshift range $0.04 \leq z \leq 0.14$ ($z_{median}\simeq0.1$) and mass range $10.3 \leq \log(M_\ast/M_\odot) \leq 11.5$. We further require a redshift-dependent lower mass limit, $\log(M_\ast/M_\odot) \geq 10.6 + 2.28 \times \log(z/0.1)$, the mass completeness limit of the SDSS spectroscopic sample \citep{cha15}.

The SDSS spectra are obtained with a fiber spectrograph, while the LEGA-C spectra is obtained with slits. To make a proper comparison between the two datasets, we first require a $g$-band fiber aperture covering fraction of $\geq 20\%$ from the comparison of the 3-arcsecond fiber flux with the total flux to mitigate the bias that fiber spectra sample only the central part of galaxies. 
We then apply a statistical correction on the D$_n$4000 and EW(H$\delta$) to account for the age gradients of galaxies. We describe the derivation of the statistical correction in Section~\ref{sec:corr}.

We adopt the stellar mass and spectral measurements by the MPA/JHU group \citep{kau03a,bri04,sal07}. The stellar masses are estimated by SED fitting, using templates constructed from the \citet{bc03} population synthesis code, assuming a range of star-formation histories and metallicities with a \citet{cha03} IMF. The basic assumptions are the same as the templates we used for deriving the stellar masses of LEGA-C galaxies.

For the D$_n$4000 and EW(H$\delta$), we adopt the measurement on the data after subtracting emission lines.
To account for volume incompleteness, each galaxy is assigned a weight $1/V_{max}$, where $V_{max}$ is the maximum volume for which the galaxy would be included in the sample based on our redshift-dependent lower mass limit. The redshift and mass distributions of the SDSS sample are shown in Fig.~\ref{fig:mzs}.

\subsection{Estimating the bias on indices introduced by SDSS fibers}
\label{sec:corr}

The SDSS fiber spectra probe the central part of galaxies. Recent IFU surveys have shown that galaxies in the local Universe have on average negative age gradients, i.e., galaxy outskirts are younger than galaxy center \citep{gon15,god17,wan17}. Any redshift evolution is therefore exacerbated if we use SDSS fiber spectra to create a low-redshift baseline sample, as those measurements will be biased toward old ages. In the local Universe, age gradients depend on galaxy morphological types, where early-type galaxies have only mild age gradients but late-type galaxies, especially Sa and Sb galaxies, exhibit strong age gradients \citep{gon15,god17}. Estimating the aperture bias by galaxy types is thus necessary.

\citet{wan17} measured the D$_n$4000 and EW(H$\delta$) as a function of the effective radius ($R_e$) out to $1.5 R_e$ for galaxies in the Mapping Nearby Galaxies at APO \citep[MaNGA,][]{bun15} survey. They reported the profiles of D$_n$4000 and EW(H$\delta$) as a function of stellar masses and star-formation properties of galaxies \citep[Fig.~8 in ][]{wan17}.  

Briefly, \citet{wan17} presented the index gradients of three types of galaxies, categorized by the equivalent width of H$\alpha$ emission and D$_n$4000: 'star-forming', 'partially quenched', and 'totally quenched'. The radial profiles of indices of 'star-forming' and 'partially quenched' galaxies are similar, therefore, we take the average of the two and refer them as 'star-forming' hereafter.

We use the index gradients to derive a statistical correction for our SDSS comparison sample. Using the slopes of D$_n$4000 and EW(H$\delta$) presented in Figure~8 of \citet{wan17}, we calculate the difference between indices measured from the integrated light within $0.5 R_e$ and $1.5 R_e$ as the correction to be applied to the SDSS fiber measurements. Assuming a S\'{e}rsic $n=4$ light profile, the two radii enclose $\sim30\%$ and $\sim60\%$ of total light, similar to the median fiber and slit covering fraction of our SDSS and LEGA-C sample, respectively. 

We apply the correction of 'totally quenched' galaxies to galaxies with weak H$\alpha$ emission ($\mbox{EW}(\mbox{H}\alpha)>-1$\AA), and the correction of 'star-forming' galaxies to the rest. This scheme is motivated by Fig.~11 of \citet{wan17}, which showed that the integrated EW(H$\alpha$) within 0.5$R_e$ serves as a reasonable demarcation between the two types of galaxies. 

\begin{figure}
\includegraphics[width=\columnwidth]{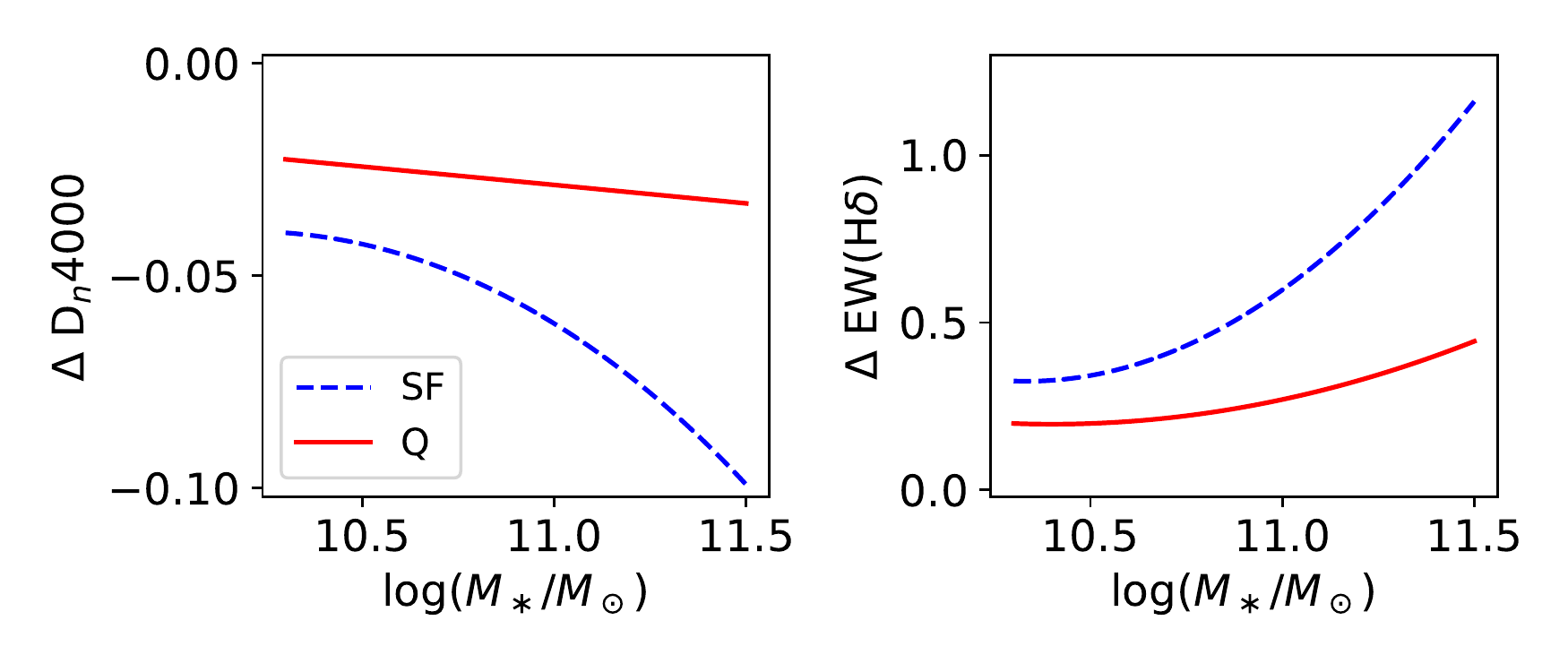}
\caption{The statistical correction on D$_n$4000 and EW(H$\delta$) for the SDSS sample. The correction depends on the stellar mass and the equivalent width of the H$\alpha$ emission line (see Section~\ref{sec:data}). The curve shows the values to be added to indices measured from the SDSS fiber spectra.}
\label{fig:corr}
\end{figure}

In summary, the correction to the SDSS sample depends on stellar mass and the equivalent width of H$\alpha$ emission in the fiber (Fig.~\ref{fig:corr}). The correction is larger for 'star-forming' galaxies than 'quiescent' galaxies, qualitatively consistent with the expectation from galaxy morphological types \citep{gon15,god17}.
We implicitly assume that all SDSS galaxies have a S\'{e}rsic $n=4$ light profile and the fiber covers out to $0.5 R_e$ then correct the indices to the values as they were observed out to $1.5 R_e$. 
Different S\'{e}rsic profiles have little effect; the correction differs by $\sim20\%$ between $n=1$ and $n=6$. A more accurate comparison would involve creating mock slit spectra from MaNGA or other local IFU surveys like CALIFA \citep{san12,wal14} and SAMI \citep{bry15}, mimicking the observing condition and aperture size of the LEGA-C survey (Bezanson et al. 2017, submitted). 

The corrected D$_n$4000 is smaller and EW(H$\delta$) is larger than the measured values (Fig.~\ref{fig:corr}). Galaxies shift along the locus on the D$_n$4000-EW(H$\delta$) plane in Section~\ref{sec:result}.  We also repeat the analysis in this paper using uncorrected indices. The inferred stellar age in Section~\ref{sec:age} becomes $<1$~Gyr older. Our main conclusion in the paper does not change. 

\section{The 4000~\AA~break and Balmer absorption strength of galaxies at $z\sim0.8$}
\label{sec:result}

About half of stars in the present-day Universe formed since $z\sim1$ \citep{dic03,rud03,ilb10,muz13}. The stellar population at $z\sim0.8$ is thus expected to be very different from galaxies in the local Universe. With over 1000 high-quality spectra, we are able to construct the distributions of D$_n$4000, EW(H$\delta$), and for the first time, the distribution of galaxies on the D$_n$4000--EW(H$\delta$) plane at $\sim7$~Gyrs look-back time. In this section, we present the inventories of stars in galaxies of the same stellar masses at two epochs.


\subsection{The distribution of \dn\ and \EWHd\ as a function of stellar masses}

\begin{figure*}
	\includegraphics[width=\textwidth]{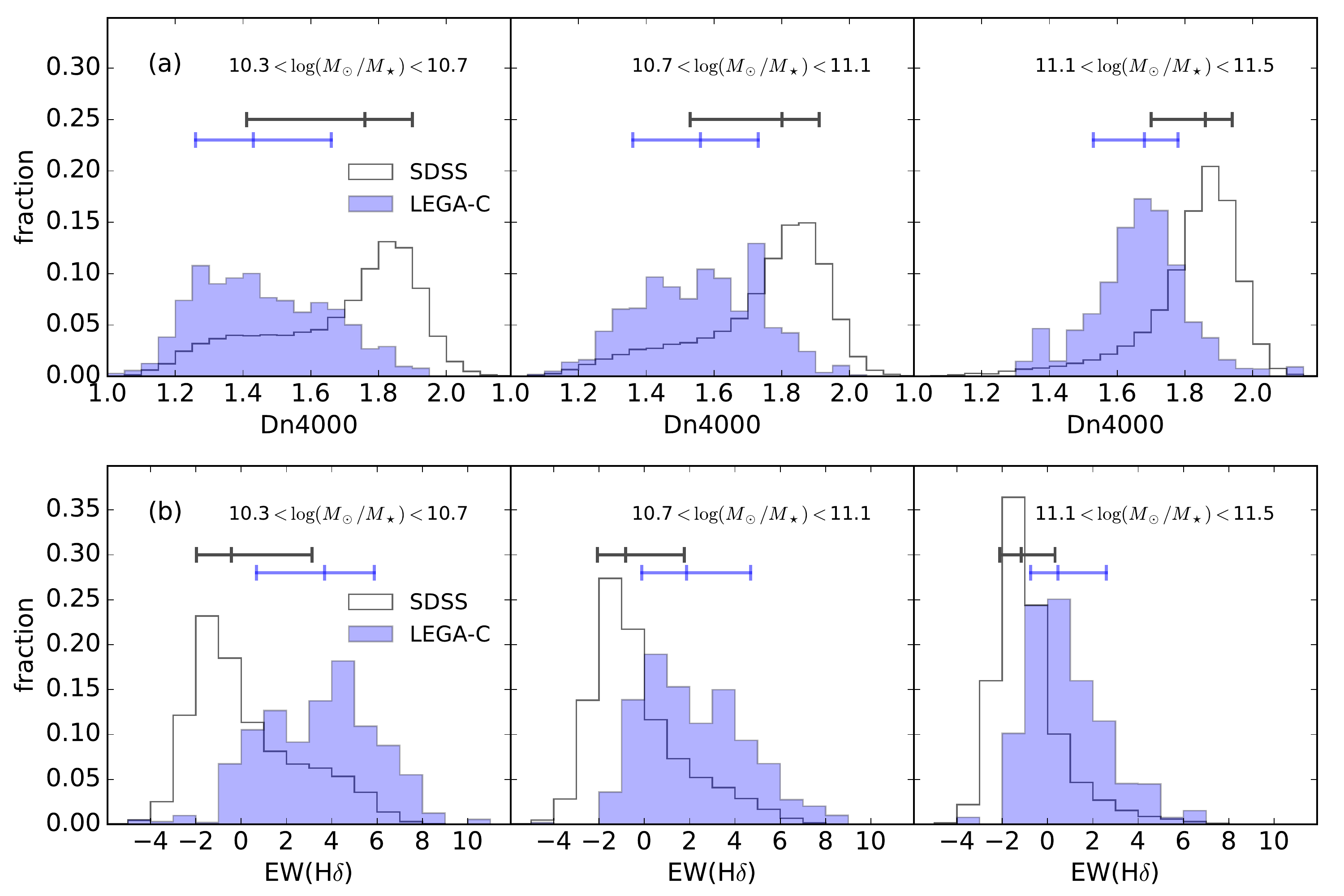}
	\caption{Distribution of D$_n$4000 and EW(H$\delta$) of LEGA-C (blue, $z\sim0.8$) and SDSS (white, $z\sim0.1$) samples with completeness correction. Each panel shows galaxies in 0.4~dex stellar mass bins. The errorbars indicate the 16th, 50th, and 84th percentiles of the distributions. At fixed stellar mass, LEGA-C galaxies have on average smaller D$_n$4000 and larger EW(H$\delta$), indicating younger populations. At $z\sim0.8$, the distributions of both D$_n$4000 and EW(H$\delta$) depend on stellar mass.}
	\label{fig:DnM}
\end{figure*}

Fig.~\ref{fig:DnM} shows the histogram of D$_n$4000 and EW(H$\delta$) of the completeness-corrected LEGA-C and SDSS samples in each stellar mass bin. The median, 68th, and 95th percentiles of the distribution are listed in Table~\ref{tbl:dist}. 

\begin{table*}
	\caption{D$_n$4000 and EW(H$\delta$) Distributions as a Function of Stellar Mass}
	\begin{center}
		\begin{tabular}{cccccccccccc}
			\hline
			\hline
			& \multicolumn{11}{c}{LEGA-C, $z \sim 0.8$} \\
			\cline{2-12}
			$\log(M_\ast/M_\odot)$ & \multicolumn{5}{c}{D$_n$4000} & & \multicolumn{5}{c}{EW(H$\delta$)} \\
			\cline{2-6}  \cline{8-12}
			& 2.5\% & 16\% & 50\% & 84\% & 97.5\% & & 2.5\% & 16\% & 50\% & 84\% & 97.5\% \\
			\hline
			$10.3 < \log(M_\ast/M_\odot)<10.7$ & 1.16 & 1.26 & 1.43 & 1.66 & 1.83 & & -0.93 & 0.68 & 3.69 & 5.88 & 7.86 \\
			$10.7 < \log(M_\ast/M_\odot)<11.1$ & 1.21 & 1.36 & 1.56 & 1.73 & 1.88 & & -1.19 & -0.12 & 1.86 & 4.70 & 7.25 \\
			$11.1 < \log(M_\ast/M_\odot)<11.5$ & 1.35 & 1.53 & 1.68 & 1.78 & 1.94 & & -1.49 & -0.74 & 0.45 & 2.60 & 5.66 \\
			\hline
			All & 1.17 & 1.30 & 1.49 & 1.71 & 1.86 & & -1.16 & 0.23 & 2.94 & 5.52 & 7.76 \\
			
			\hline
			\hline
			& \multicolumn{11}{c}{SDSS, $z \sim 0.1$} \\
			\cline{2-12}
			$\log(M_\ast/M_\odot)$ & \multicolumn{5}{c}{D$_n$4000} & & \multicolumn{5}{c}{EW(H$\delta$)} \\
			\cline{2-6}  \cline{8-12}
			& 2.5\% & 16\% & 50\% & 84\% & 97.5\% & & 2.5\% & 16\% & 50\% & 84\% & 97.5\% \\
			\hline
			$10.3 < \log(M_\ast/M_\odot)<10.7$ & 1.21 & 1.41 & 1.76 & 1.90 & 2.00 & & -3.14 & -1.97 & -0.44 & 3.12 & 5.69 \\
			$10.7 < \log(M_\ast/M_\odot)<11.1$ & 1.26 & 1.53 & 1.80 & 1.91 & 2.01 & & -3.19 & -2.07 & -0.82 & 1.77 & 5.00 \\
			$11.1 < \log(M_\ast/M_\odot)<11.5$ & 1.39 & 1.70 & 1.86 & 1.94 & 2.02 & & -3.00 & -2.11 & -1.17 & 0.33 & 3.48 \\
			\hline
			All & 1.23 & 1.46 & 1.78 & 1.91 & 2.00 & & -3.15 & -2.02 & -0.66 & 2.55 & 5.47 \\ 
			\hline

		\end{tabular}
	\end{center}
	\label{tbl:dist}
\end{table*}

At $z\sim0.8$, the D$_n$4000 distribution depends on the stellar mass. The median D$_n$4000 shifts from 1.43 in the low mass bin to 1.68 in the high mass bin. The distribution of D$_n$4000 is narrower in the high mass bin, as quantified by the 16th and 84th percentiles. There is only a small fraction of galaxies with $\mbox{D}_n4000 < 1.4$, which is the median value of the low mass bin. This result is in broad agreement with the distribution measured from the VVDS and the VIPERS survey based on lower S/N spectra \citep{ver08,hai17}. At $z\sim0.1$, the D$_n$4000 distribution depends less on mass, with peaks at $\mbox{D}_n4000 \simeq 1.8$ at all masses. The major difference is that the tail to low D$_n$4000 vanishes, as can be seen from the 2.5 and 16 percentiles in Table~\ref{tbl:dist}.

Fig.~\ref{fig:DnM}b shows for the first time the distribution of EW(H$\delta$) at $z\sim0.8$. Similar to the distribution of D$_n$4000, the EW(H$\delta$) distribution at $z\sim0.8$ also depends strongly on the stellar mass. In the low mass bin, the EW(H$\delta$) distributes around $\mbox{EW}(\mbox{H}\delta) \simeq 4$\AA. In the high mass bin, the median shifts to $\mbox{EW}(\mbox{H}\delta) \simeq 0$\AA\ and there are very few galaxies with $\mbox{EW}(\mbox{H}\delta) > 4$\AA. On the other hand, the distributions at $z\sim0.1$ center at $\mbox{EW}(\mbox{H}\delta) \simeq -1$\AA\ for all masses. Similarly, the tail to the younger end (larger EW(H$\delta$)) vanishes in the high mass bin.

\subsection{The D$_n$4000--EW(H$\delta$) plane}
\label{sec:DnHd}

\begin{figure*}
	\centering
	\includegraphics[width=\textwidth]{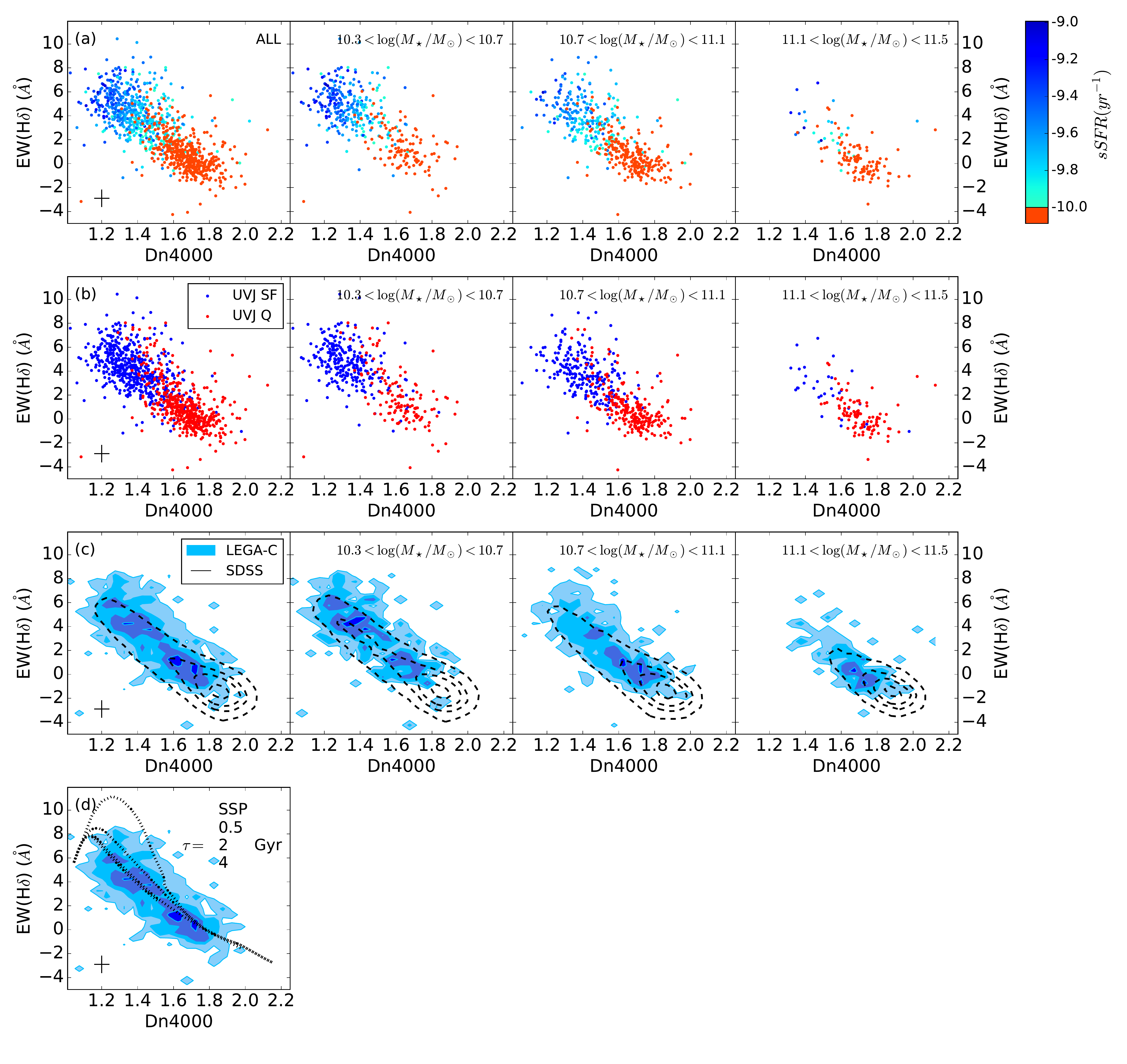}
	\caption{The distribution of LEGA-C and SDSS galaxies on the D$_n$4000--EW(H$\delta$) plane in different stellar mass bins. (a) The colors dots are individual LEGA-C galaxies, color-coded by sSFR. Galaxies with $sSFR < 10^{-10}$~yr$^{-1}$ are in red. The D$_n$4000 correlates with sSFR, where high sSFR galaxies have small D$_n$4000. The cross in the bottom-left corner is the typical uncertainty. The EW(H$\delta$) uncertainty is smaller than the EW(H$\delta$) distribution at D$_n$4000, therefore, our measurements resolve the recent star-formation histories of individual galaxies through EW(H$\delta$). (b) The same LEGA-C galaxies as in panel (a), color-coded by the star-forming/quiescent classification in the UVJ two-color scheme. The star-forming galaxies and quiescent galaxies are roughly separated by $\mbox{D}_n4000 \simeq 1.55$ and $\mbox{EW}(\mbox{H}\delta) \simeq 2$\AA. (c) Distributions of completeness-corrected LEGA-C and SDSS samples. Blue filled contours represent the LEGA-C sample and the dashed contours represent the SDSS sample. Contours levels are at the 0.05, 0.20, 0.40, and 0.80 times the peak value for each sample. The LEGA-C sample exhibit a bimodal distribution on the D$_n$4000--EW(H$\delta$) plane, while the SDSS sample does not. (d) An illustration of how galaxy evolves on the D$_n$4000--EW(H$\delta$) plane with different SFH. Four SFH are shown (top to bottom): SSP, 0.5, 2, and 4 Gyr $\tau$ decaying time. All models are with solar metallicity. The models with 2 and 4 Gyr $\tau$ decaying time occupy almost the same loci. The contour levels are the same as in panel (c).}
	\label{fig:HD4000}
\end{figure*}

Fig.~\ref{fig:HD4000} shows LEGA-C galaxies on the D$_n$4000--EW(H$\delta$) plane. Overall, galaxies at $z\sim0.8$ are located along a diagonal sequence on the D$_n$4000--EW(H$\delta$) plane. As the stellar mass increases, the population moves from the top-left towards the bottom-right corner of the panel, i.e., larger $\mbox{D}_n4000$ and smaller EW(H$\delta$), indicating an overall older stellar population in more massive galaxies \citep{kau03b,siu17}. 

In Fig.~\ref{fig:HD4000}a, galaxies are color-coded according to the specific star-formation rate (sSFR), the SFR divided by the stellar mass. The sSFR and $\mbox{D}_n4000$ are correlated such that galaxies with high sSFRs also have small $\mbox{D}_n4000$. The correlation between the sSFR and the $\mbox{D}_n4000$ is qualitatively similar to the correlation found for galaxies at $z\sim0.1$ \citep{bri04}. 

Fig.~\ref{fig:HD4000}b shows again the LEGA-C galaxies. Star-forming galaxies and quiescent galaxies in the UVJ two-color classification scheme \citep{muz13} are plotted in blue and red, respectively. The star-forming/quiescent classification based on the UVJ colors and sSFR, adopting $sSFR=10^{-10}$~ yr$^{-1}$ as demarcation, are in good agreement. On the D$_n$4000--EW(H$\delta$) plane, the star-forming and quiescent galaxies can be roughly separated by  $\mbox{D}_n4000\simeq1.55$ and/or $\mbox{EW}(\mbox{H}\delta)\simeq2$\AA. 

Fig.~\ref{fig:HD4000}c shows the density contours of LEGA-C galaxies with the completeness correction in blue and SDSS galaxies in black. For galaxies with $10.3 < \log(M_\ast/M_\odot) < 11.5$, the LEGA-C distribution is double-peaked, with a valley located at $\mbox{D}_n4000\simeq1.55$ and $\mbox{EW}(\mbox{H}\delta)\simeq2$\AA, corresponding to the separation between star-forming and quiescent galaxies. This bimodal distribution of galaxies on the D$_n$4000--EW(H$\delta$) plane is also present in the nearby Universe with similar demarcation \citep{kau03b}. 

Galaxies at $z\sim0.8$ and $z\sim0.1$ occupy a qualitatively similar locus on the D$_n$4000-EW(H$\delta$) plane but populate this locus differently. At $z\sim0.1$, the distribution peaks at $\mbox{D}_n4000 \sim 1.9$ and $\mbox{EW}(\mbox{H}\delta) \sim -2$\AA. On the contrary, quiescent galaxies at $z\sim0.8$ have on average smaller D$_n$4000 and larger EW(H$\delta$). Also, there are very few galaxies at $z\sim0.8$ with $\mbox{D}_n4000 > 1.9$ or $\mbox{EW}(\mbox{H}\delta) < -2$\AA. Furthermore, Fig.~\ref{fig:HD4000}c shows that the distribution of LEGA-C galaxies extends to higher EW(H$\delta$), especially for galaxies with small D$_n$4000. Previous studies based on Principal Component Analysis of spectra also suggest a higher fraction of galaxies with strong H$\delta$ at higher redshifts \citep{wil09,row18}.

\section{The strong H$\delta$ absorption at $z\sim0.8$}
\label{sec:hd}

\begin{figure*}
	\includegraphics[width=\textwidth]{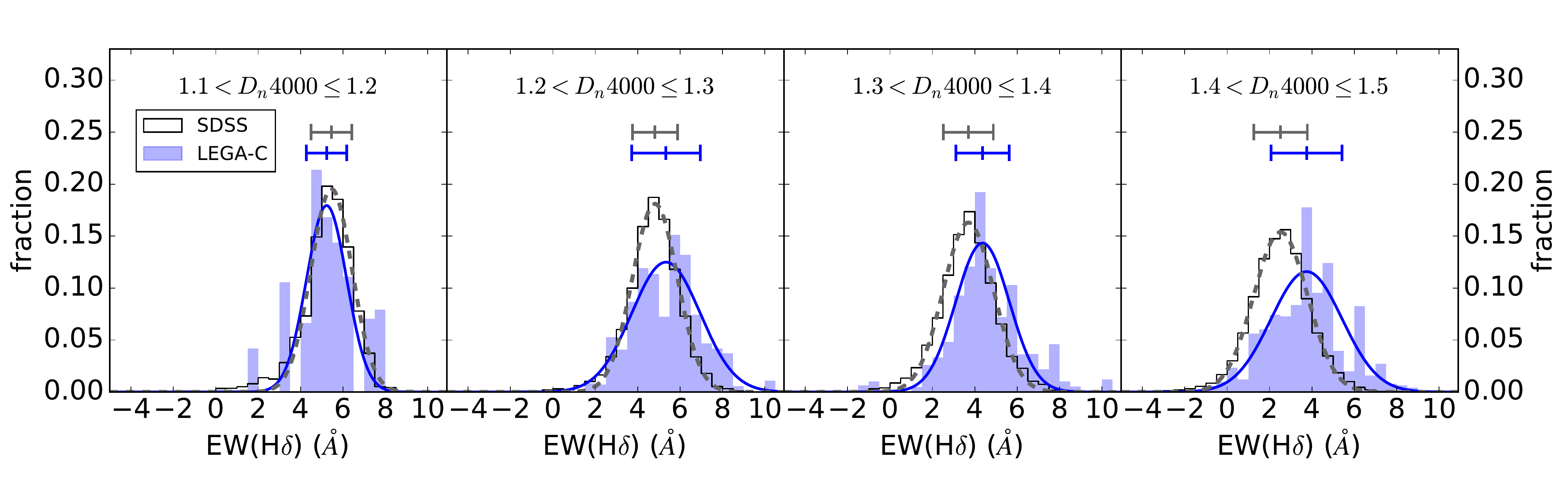}
	\caption{The distribution of EW(H$\delta$) at fixed D$_n$4000. The white and blue histograms show the distribution of completeness-corrected SDSS and LEGA-C galaxies, respectively. The gray dashed line and the blue solid lines are the best-fit Gaussian to each histogram. The central EW(H$\delta$) and the dispersion of the best-fit Gaussians are labeled as the errorbars. The best-fit Gaussian parameters are listed in Table~\ref{tab:hd}. At $\mbox{D}_n4000>1.2$, more galaxies at $z\sim0.8$ have large EW(H$\delta$) and the distribution is also wider. The difference between the EW(H$\delta$) distribution implies that the SFR of star-forming galaxies at $z\sim0.8$ changes more rapidly than star-forming galaxies at $z\sim0.1$. }
	\label{fig:hd}
\end{figure*}

Fig.~\ref{fig:hd} shows the distribution of EW(H$\delta$) in four narrow D$_n$4000 bins for galaxies with $\mbox{D}_n4000 \leq 1.5$, where star-forming galaxies dominate the population. Except for the lowest D$_n$4000 bin, the EW(H$\delta$) distributions of galaxies $z\sim0.8$ extend to larger EW(H$\delta$) and are on average broader. We fit a Gaussian profile to each EW(H$\delta$) distribution and list the best-fit parameters in Table~\ref{tab:hd}. 

The strong Balmer absorption lines in star-forming galaxies are usually interpreted as evidence for a rapidly declining star-formation rate in the last $\lesssim 1$~Gyr. 
An illustration is shown in Fig.~\ref{fig:HD4000}d. We plot \citet{bc03} evolutionary tracks of 4 different star-formation histories with solar metallicity: an SSP and 3 exponential-decay SFHs with 0.5, 2, and 4 Gyr decaying time $\tau$. 
The strength of the H$\delta$ absorption increases after the O- and B-type stars fade away and the A-type stars dominate the optical spectrum.
Rapidly declining SFHs, e.g., SSP or $\tau=0.5$~Gyr, will elevate the EW(H$\delta$) at $\mbox{D}_n4000 \lesssim 1.5$ for several hundred Myrs comparing to a more gently declining SFH. Thus, the higher EW(H$\delta$) suggests that the SFRs of $z\sim0.8$ star-forming galaxies change more rapidly than in low-$z$ star-forming galaxies. 

Based on observed evolution of the star-formation main sequence (MS), \citet{lei12} derived analytic formulae for average SFHs of star-forming galaxies. We can thus calculate the average declining rate of the SFRs in the 1~Gyr period prior to $z\sim0.8$ and $z\sim0.1$. Adopting the parametrized MS evolution $\psi(M_\ast,z) \propto M_\ast^{1+\beta}(1+z)^\alpha$ with $\alpha=3.45$ and $\beta=-0.35$ \citep[][see also \citet{dam09,oli10,fum12}]{kar11} and the analytic formulae in \citet{lei12}, the average SFHs of star-forming galaxies in the 1~Gyr period prior to $z\sim0.8$ and $z\sim0.1$ can be approximated by the $\tau$ model with $\tau\simeq2$~Gyrs and $\tau\simeq4$~Gyrs, respectively. 

Exponentially declining SFH models with $\tau\simeq2$~Gyr and $\tau\simeq4$~Gyr occupy very similar loci on the D$_n$4000--EW(H$\delta$) plane, therefore, the increase in average SFR from $z\sim0.1$ to $z\sim0.8$ does not explain the stronger H$\delta$ absorption at $z\sim0.8$. Instead, the strong H$\delta$ absorption implies that the SFRs of individual galaxies have stronger time variabilities than the average evolution of the star-formation MS at $z\sim0.8$. A star-forming galaxy may experience starburst events while it stays in the MS or oscillate up and down within the MS in a timescale shorter than the evolution of average sSFR. 
Galaxies with recent rapidly declining SFHs will have stronger Balmer absorptions and deviate from the main locus on the D$_n$4000-EW(H$\delta$) plane for a few hundred Myrs, create an excess at large EW(H$\delta$) and the EW(H$\delta$) distribution becomes broader \citep{kau03a}. 

The high variability SFRs in star-formation galaxies at higher redshifts is also suggested by the cosmological zoom-in simulations. The Feedback in Realistic Environments \citep[FIRE;][]{hop14} showed that all galaxies at high redshifts ($z\gtrsim1$) have bursty SFHs, while massive, $\sim L_\ast$ galaxies settle to steady SFHs at $z\lesssim1$ \citep{spa17,orr17,fau18}. The strong time variability of SFRs have been observed in local dwarf galaxies by comparing SFRs derived from H$\alpha$ and FUV emission, which trace different timescales \citep{wei12}. The large scatter of the EW(H$\delta$) in low-mass galaxies at $z\sim0.1$ is another sign of burty SFHs \citep{kau03a,kau14}. At $z\sim0.7$, \cite{guo16} used H$\beta$ and FUV and found that the SFRs of low-mass galaxies ($M_\ast < 10^{9.5} M_\odot$) have stronger time variability than galaxies at low redshifts. This redshift evolution is in qualitative agreement with numerical simulations.  
Our result provides an evidence that the SFRs of higher mass galaxies at $z\sim0.8$ also vary at a short timescale. The rapidly changing SFRs left imprints on the stellar population through the H$\delta$ absorption, which lasts for a longer timescale of a few hundred Myrs and the difference between $z\sim0.8$ and $z\sim0.1$ is visible on the D$_n$4000-EW(H$\delta$) plane.

\begin{table*}
	\caption{Best-fit Gaussian Parameters for the EW(H$\delta$) distribution}
	\begin{center}
		\begin{tabular}{cccccccc}
			\hline
			\hline
			& \multicolumn{3}{c}{LEGA-C, $z\sim0.8$} & & \multicolumn{3}{c}{SDSS, $z\sim0.1$} \\
			\cline{2-4} \cline{6-8}
			& EW(H$\delta)_0$ & $\sigma(H\delta)$ & A & & EW(H$\delta)_0$ & $\sigma(H\delta)$ & A \\
			\hline
			$1.1 < D_n4000 \leq 1.2$ & 5.23$\pm$0.13 & 0.97$\pm$0.13 & 0.17$\pm$0.02 & & 5.46$\pm$0.02 & 0.97$\pm$0.02 & 0.19$\pm$0.00 \\
			$1.2 < D_n4000 \leq 1.3$ & 5.34$\pm$0.11 & 1.62$\pm$0.11 & 0.12$\pm$0.01 & & 4.81$\pm$0.02 & 1.06$\pm$0.02 & 0.18$\pm$0.00 \\
			$1.3 < D_n4000 \leq 1.4$ & 4.36$\pm$0.10 & 1.26$\pm$0.10 & 0.14$\pm$0.01 & & 3.70$\pm$0.02 & 1.18$\pm$0.02 & 0.16$\pm$0.00 \\
			$1.4 < D_n4000 \leq 1.5$ & 3.73$\pm$0.17 & 1.68$\pm$0.17 & 0.12$\pm$0.01 & & 2.52$\pm$0.01 & 1.27$\pm$0.01 & 0.15$\pm$0.00 \\
			\hline
		\end{tabular}
		\begin{tablenotes}
			\item[] The Gaussian model is $A \times \exp[ -(EW(H\delta)-EW(H\delta)_0 )^2/(2\times\sigma(H\delta)^2) ]$
		\end{tablenotes}
	\end{center}
	\label{tab:hd}
\end{table*}

Except for SFHs, the spectral indices also depend on the stellar metallicity and are affected by dust extinction. Based on the stellar mass-stellar metallicity relation presented by \citet{gal14}, a solar metallicity is in general a good approximation for both $z\sim0.8$ and $z\sim0.1$ populations. Only galaxies $M_\ast \lesssim 10^{10.5} M_\odot$ at $z\sim0.8$ appear to be slightly sub-solar, with an average $\log(Z_\ast/Z_\odot)=-0.21$ \citep{gal14}. We have compared the loci of the \citet{bc03} models of solar and sub-solar metallicity ($Z_\ast/Z_\odot = 0.4$) with various SFHs on the D$_n$4000-EW(H$\delta$) plane. We find that the sub-solar metallicity does not produce larger EW(H$\delta$) at fixed D$_n$4000. 

Alternatively, dust can alter both the D$_n$4000 and EW(H$\delta$). The D$_n$4000, which is essentially a color index, will in general be larger when the dust attenuation is more severe \citep{mac05}. The effect of dust on the EW(H$\delta$) depends on the dust geometry. The EW(H$\delta$) will be boosted up if the dust is distributed mainly around the birth clouds of young stars. In this case, the featureless continuum of hot stars is obscured and the Balmer absorption feature from intermediate age stars becomes more prominent. On the other hand, the diffuse interstellar dust has little effect on the measured EW(H$\delta$) \citep{mac05}. 

If the difference in the EW(H$\delta$) distribution is entirely due to the dust attenuation, galaxies at $z\sim0.8$ must have a birth cloud V-band attenuation $A_V\simeq2$ magnitudes larger than that of SDSS galaxies to elevate the EW(H$\delta$) by $\sim$1\AA\ \citep{mac05}. On the other hand, if we artificially decrease the D$_n$4000 of LEGA-C galaxies by $\sim0.07$, the EW(H$\delta$) distributions at fixed D$_n$4000 match that of the SDSS galaxies better. This shift in D$_n$4000 indicates that LEGA-C galaxies have $A_V$ more than 1.5 magnitudes larger than SDSS galaxies, assuming the \citet{car89} extinction law. In either case, such a heavy extinction is inconsistent with previous studies, which found that the dust extinction of star-forming galaxies at $z\sim0.8$ is similar to or only slightly higher than galaxies at $z\sim0.1$ of the same stellar mass \citep[][see also \citet{sob12,dom13,kas13} for results up to $z\sim1.6$]{gar10a,gar10b,zah13,les18}.

In summary, the large EW(H$\delta$) can only be explained by a rapidly changing SFR at $z\sim0.8$. Changes in metallicity and dust attenuation cannot explain it. A full analysis incorporating star-formation history, metallicity, and dust requires using more spectral features, i.e., full-spectral fitting and/or combing with multi-wavelength photometry \citep[e.g., ][]{pac12,pac13}. We will present the star-formation histories of individual galaxies at $z\sim0.8$ constructed from the LEGA-C spectra in forthcoming papers (Chauke et al., submitted; Pacifici et al. in prep.).

\section{Stellar ages from D$_n$4000 and EW(H$\delta$)}
\label{sec:age}

\begin{figure*}
	\includegraphics[width=\textwidth]{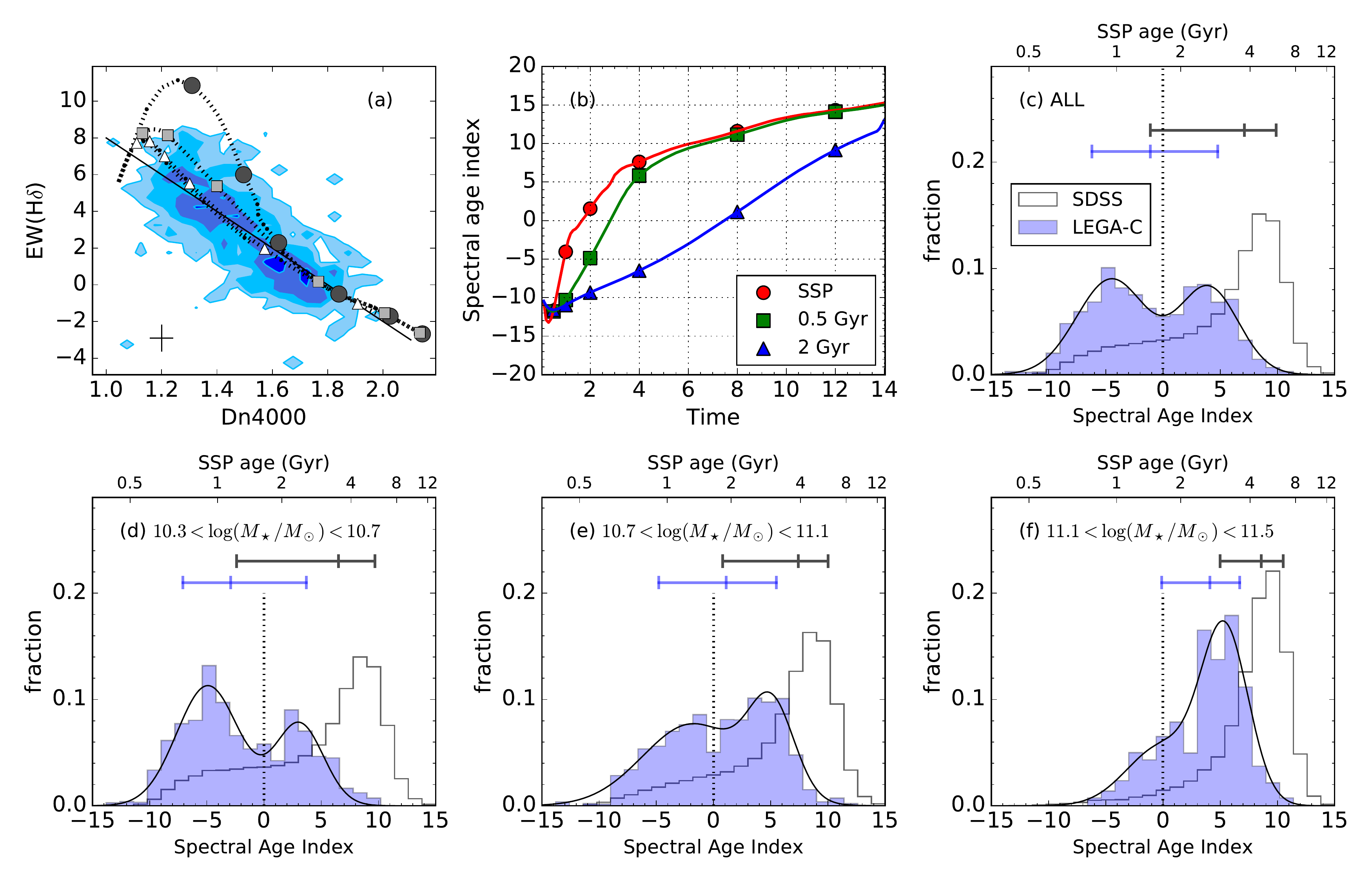}
	\caption{(a) Distribution of LEGA-C galaxies on the D$_n$4000--EW(H$\delta$) plane, overplotted with model evolutionary tracks. The contours are the same as in Fig.~\ref{fig:HD4000}c and model tracks are the same as in Fig.~\ref{fig:HD4000}d. The $\tau=2$~Gyr and the $\tau=4$~Gyr models overlap with each other. Time steps of 0.5, 1, 2, 4, 8, and 12 Gyrs are marked with black circle, gray squares, and white triangles for the SSP, $\tau=0.5$~Gyr, and $\tau=2$~Gyr models, respectively. The black line indicates the ridge line of the distribution. (b) The spectral age index as a function of time with different SFHs. The definition of the age spectral index is explained in Section~\ref{sec:age}. The red, green, and blue curves represent for SSP, $\tau=0.5$~Gyr, and $\tau=2$~Gyr SFHs, respectively. The same time steps as in panel (a) are labeled. (c,d,e,f) The distribution of the spectral age index of LEGA-C galaxies (blue) and SDSS galaxies (white). Galaxies with older stellar populations have larger indices. The SSP ages are labeled on the top of each panel, assuming a solar metallicity. The solid curves are the best-fit two-Gaussian models of the LEGA-C sample. The errorbars indicate the 16th, 50th, and 84th percentiles of the distributions. }
	\label{fig:project}
\end{figure*}

The D$_n$4000 and EW(H$\delta$) are commonly used as proxies for the light-weighted stellar ages. Fig.~\ref{fig:project}a shows again the distribution of LEGA-C galaxies on the D$_n$4000-EW(H$\delta$) plane, together with \citet{bc03} evolutionary tracks of SSP and exponential-decline SFHs with $\tau=$0.5, 2, and 4 Gyr with solar metallicity. 
For the stellar mass range discussed in this paper, a solar metallicity population is a good approximation for galaxies at both $z\sim0.1$ and $z\sim0.8$ \citep{gal14}.

Motivated by the evolutionary tracks in Fig.~\ref{fig:project}a, we combine D$_n$4000 and EW(H$\delta$) to construct the distribution of galaxies along the ridge line of the diagonal distribution on the D$_n$4000--EW(H$\delta$) plane. We compute a spectral age index as $15\times \mbox{D}_n4000 - \mbox{EW}(\mbox{H}\delta) - 20.5$. This new index represents for the distribution on the D$_n$4000--EW(H$\delta$) plane projected onto the ridge of LEGA-C contours (the black line in Fig.~\ref{fig:project}a). The ridge of LEGA-C contours tracks closely to the $\tau=2$~Gyr model as well as the SSP model for old populations. If galaxies evolve as the model SFHs, galaxies of the same age have the same spectral age index. A larger index corresponds to an older stellar population. The constant $-20.5$ is chosen so that the zero-point falls between the bimodal distribution \citep[e.g., $\mbox{D}_n4000 \simeq 1.55$ and $\mbox{EW}(\mbox{H}\delta)\simeq2$\AA;][]{kau03b,hai17}. The spectral age indices can be translated into ages according to Fig.~\ref{fig:project}b based on different SFHs. 

Fig.~\ref{fig:project}c shows the distributions of the spectral age indices of the SDSS and the LEGA-C sample. Fig.~\ref{fig:project}d,e, and f show the distributions in each stellar mass bins. The corresponding SSP ages are labeled on the top of the panels. The median, 68th, and 95th percentiles of the distribution are listed in Table~\ref{tbl:ind}. 
We note that the SSP ages should be interpreted with care. For star-forming galaxies, a single number of a luminosity-weighted age may not be a good quantitative age diagnostic \citep[][Chauke et al., submitted]{zib17}. On the other hand, for very old stellar populations, the spectral indices evolve little with time (see Fig.~\ref{fig:project}b), thus, not sensitive to stellar ages. Also, we assume a solar metallicity for all galaxies. The age would be underestimated if galaxies have sub-solar metallicities and vise versa. 
The spectral age indices and the corresponding SSP ages are slightly affected by the dust extinction. Assuming a typical extinction at $z\sim1$, we estimate a $<0.5$~Gyr effect on the SSP ages for both star-forming and quiescent galaxies. For comparing the age difference between $z\sim0.8$ and $z\sim0.1$, the effect of dust is likely minimum because of the similar amount of extinction in both populations \citep{sob12,dom13,kas13,gal14}.

At $z\sim0.8$, the age increases with stellar mass. The mass-dependent stellar ages supports the downsizing galaxy formation, where more massive galaxies formed in earlier times \citep{tho10} and this archaeological trend is already in place in the first half of the cosmic time. The oldest galaxies with $M_\ast > 10^{11} M_\odot$ are $\sim5$~Gyr old, indicating that they form at $z\gtrsim3$. The formation redshifts are similar to those $z>3$ quiescent galaxies spectroscopically-confirmed recently \citep{gob12,str15,gla17}.

\begin{figure}
	\includegraphics[width=\columnwidth]{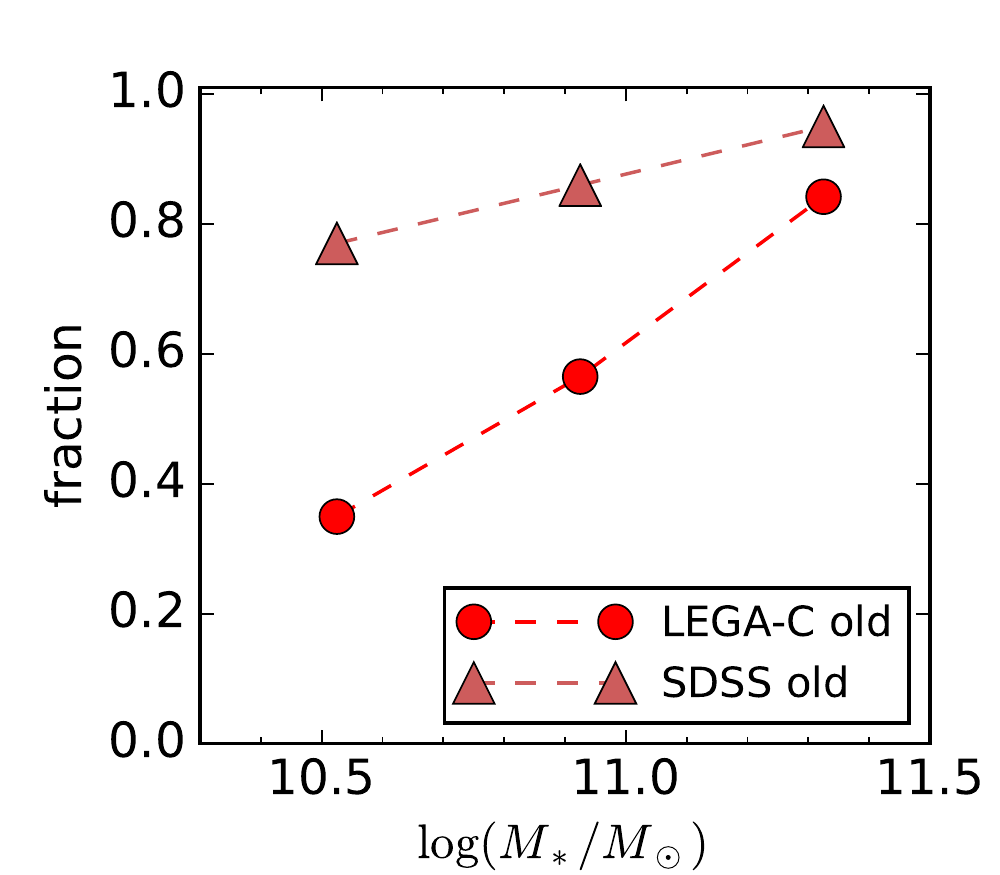}
	\caption{The relative abundance of the old galaxy populations as a function of stellar masses. The LEGA-C samples are shown by filled circles. The SDSS samples are shown by filled triangles. The uncertainties are smaller than the symbols. At $z\sim0.8$, the relative abundance of the old population depends on the stellar mass, from $\lesssim40\%$ to $\gtrsim80\%$ among three mass bins. On the other hand, at $z\sim0.1$, galaxies belong mostly to the old population at all masses discussed in this paper.}
	\label{fig:sfq}
\end{figure}

The distribution of the spectral age indices of the LEGA-C sample is double-peaked: the distribution of the spectral age index is better fit by a 2-Gaussian model than a single Gaussian model. 
Using the F-test, we find that for the entire population and the two lowest mass bins, the null hypothesis that an 1-Gaussian model provides no better fit than a 2-Gaussian model is rejected with probability of 0.01. On the other hand, the highest mass bin does not show as a clear bimodal distribution as in other mass bins. 
The overall bimodal spectral age indice distribution implies a bimodal light-weighted stellar age distribution. Fig.~\ref{fig:sfq} shows the fraction of galaxies with old stellar population ($\mbox{spectral age index}>0$) as a function of stellar masses. 
At $z\sim0.8$, the fraction of the old population changes sharply with the stellar mass, from $\lesssim40\%$ at the lowest mass bins to $>80\%$ at the highest mass bin. At $\log(M_\ast/M_\odot) \simeq 10.8$, the old and the young population have similar number densities. This result is in broad agreement with several previous studies which classify galaxies using either broadband colors \citep{bun06,poz10,dav13,muz13} or D$_n$4000 \citep{ver08,hai17}. 

Using the SSP age inferred from the spectral age index, we find that the massive galaxies ($\log(M_\ast/M_\odot)>11.1$) at $z\sim0.8$ are on average $\sim3$~Gyr younger than massive galaxies at $z\sim0.1$. The difference of galaxy ages is smaller than the age difference of the Universe between the two epochs ($\sim5.5$~Gyr). Pure passive evolution of the massive galaxies at $z\sim0.8$ cannot reproduce the massive galaxy population at $z\sim0.1$. 
The current analysis assumes that massive galaxies at the two epochs both have solar metallicities. The conclusion does not change if we instead use super-solar metallicities \citep{jor17}. Furthermore, if massive galaxies at lower redshifts are slightly more metal-rich, as suggested by previous studies \citep{cho14,gal14}, the inferred age difference will be even smaller, further strengthening our result. The conclusion is consistent with \citet{gal14}, who derived ages using both SSP and composite stellar populations.
Massive galaxies at high-redshifts need to acquire younger stars from either star-formation or merging with other younger galaxies. Alternatively, lower mass galaxies at $z\sim0.8$ need to grow their stellar masses and become young massive galaxies at $z\sim0.1$ \citep{bel04,gal14}. Obtaining the stellar matallicities of both star-forming and quiescent galaxies will help to constrain the evolutionary routes \citep{cho14,gal14}.

\begin{table*}
	\caption{Spectral Age Index Distributions as a Function of Stellar Mass}
	\begin{center}
		\begin{tabular}{cccccc}
			\hline
			\hline
			& \multicolumn{5}{c}{LEGA-C, $z \sim 0.8$} \\
			\cline{2-6}
			$\log(M_\ast/M_\odot)$ & \multicolumn{5}{c}{Index} \\
			\cline{2-6} 
			& 2.5\% & 16\% & 50\% & 84\% & 97.5\% \\
			\hline
			$10.3 < \log(M_\ast/M_\odot)<10.7$ & -9.3 & -7.1 & -2.9 & 3.7 & 7.2  \\
			$10.7 < \log(M_\ast/M_\odot)<11.1$ & -8.4 & -4.8 & 1.1 & 5.5 & 8.4  \\
			$11.1 < \log(M_\ast/M_\odot)<11.5$ & -5.1 & -0.1 & 4.1 & 6.7 & 9.2 \\
			\hline
			All & -9.1 & -6.2 & -1.1  & 4.8 & 7.9  \\
			\hline

			\hline
			& \multicolumn{5}{c}{SDSS, $z \sim 0.1$} \\
			\cline{2-6}
			$\log(M_\ast/M_\odot)$ & \multicolumn{5}{c}{Index}  \\
			\cline{2-6} 
			& 2.5\% & 16\% & 50\% & 84\% & 97.5\% \\
			\hline
			$10.3 < \log(M_\ast/M_\odot)<10.7$ & -7.7 & -2.4 & 6.5 & 9.7 & 11.7 \\
			$10.7 < \log(M_\ast/M_\odot)<11.1$ & -6.3 & 0.8 & 7.4 & 10.0 & 11.9 \\
			$11.1 < \log(M_\ast/M_\odot)<11.5$ & -2.8 & 5.0 & 8.6 & 10.5 & 12.1 \\
			\hline
			All & -7.3 & -1.1 & 7.1 & 9.9 & 11.8 \\
			\hline

		\end{tabular}
	\end{center}
	\label{tbl:ind}
\end{table*}

\section{Conclusion and Future Work}

We measure the D$_n$4000 and EW(H$\delta$) of 1019 galaxies at $0.6 \leq z \leq 1.0$ with $10.3 \leq \log(M_\ast/M_\odot) \leq 11.5$ using the first two years of data of the LEGA-C survey. With a typical S/N of $\sim$20 \AA$^{-1}$ and a spectral resolution $R\simeq3500$, we can separate the absorption features of the stellar continuum from the emission lines from the ISM, accurately quantifying the stellar population in both star-forming and quiescent galaxies. We show the distributions of D$_n4000$ and EW(H$\delta$) as a function of stellar mass and for the first time, where galaxies at $z\sim0.8$ are located on the D$_n$4000-EW(H$\delta$) plane for both individual galaxies and galaxies as a population. 

At $z\sim0.8$, galaxies exhibit a bimodal distribution on the $\mbox{D}_n4000$--EW(H$\delta$) plane. The star-forming and quiescent populations can be roughly separated by $\mbox{D}_n4000 = 1.55$ and $\mbox{EW}(\mbox{H}\delta) = 2$\AA\ as in the local Universe. The majority of galaxies with $\log(M_\ast/M_\odot) \lesssim 10.7$ are star-forming galaxies and populate the upper-left corner on the $\mbox{D}_n4000$--EW(H$\delta$) plane. As the stellar mass increases, galaxies have on average larger D$_n$4000 and smaller EW(H$\delta$), indicating a progressively older stellar population. 
At $\log(M_\ast/M_\odot) \gtrsim 11.1$, most galaxies have already moved onto the red sequence at $z\sim0.8$ and occupy mainly the lower-right corner on the $\mbox{D}_n4000$--EW(H$\delta$) plane. 

Using D$_n$4000 and EW(H$\delta$) as age indicators, we find that at $z\sim0.8$, more massive galaxies have older stellar populations than less massive ones, confirming the downsizing galaxy formation scenario. The oldest massive galaxies at $z\sim0.8$ are consistent with forming at $z\gtrsim3$. 

The ages of galaxies at $z\sim0.8$ and $z\sim0.1$ are inconsistent with a passive evolution scenario even for massive galaxies. Massive galaxies at $z\sim0.8$ need acquire young stars from either star-formation in galaxies and/or merging with other young galaxies, or lower mass galaxies at $z\sim0.8$ need grow masses and become younger massive galaxies at $z\sim0.1$.

At fixed D$_n$4000, star-forming galaxies at $z\sim0.8$ have on average stronger H$\delta$ absorption and the distribution of EW(H$\delta$) is wider than galaxies at $z\sim0.1$. This feature indicates that the SFR in star-forming galaxies at $z\sim0.8$ vary rapidly. The SFRs of individual galaxies change in a time scale shorter than the average evolution of the star-formation main sequence. Star-forming galaxies at $z\sim0.8$ may experience starburst events more often and/or oscillate up and down within the main sequence.

We will derive the stellar ages of individual galaxies using all available spectral features, taking into account the effects of metallicity, dust attenuation, and complex SFHs \citep{gal14}. We have carried out full spectral fitting to reconstruct the SFHs of individual galaxies (Chauke et al. 2017, submitted). These stellar age estimates of galaxies at $\sim7$~Gyr lookback time will provide new constraints on galaxy formation models. 

\acknowledgments
Based on observations made with ESO Telescopes at the La Silla Paranal Observatory under programme ID 194-A.2005 (The LEGA-C Public Spectroscopy Survey). This project has received funding from the European Research Council (ERC) under the European Union’s Horizon 2020 research and innovation programme (grant agreement No. 683184). KN and CS acknowledge support from the Deutsche Forschungsemeinschaft (GZ: WE 4755/4-1). We gratfeully acknowledge the NWO Spinoza grant.
VW acknowledges funding from the ERC (starting grant SEDmorph, PI. Wild)
JvdS is funded under Bland-Hawthorn's ARC Laureate Fellowship (FL140100278).

\bibliography{LEGAC_absline.bib}

\end{document}